\documentclass[reprint,english,aps,pre,floatfix,showpacs]{revtex4-1}

\usepackage{amsmath}
\usepackage{graphicx}
\usepackage{amssymb}
\usepackage{babel}
\usepackage{color}
\usepackage{bm}
\renewcommand{\vec}[1]{\mathbf{#1} }

\begin{document}

\title{Clustering and gelation of hard spheres induced by the Pickering effect}

\author{Andrea Fortini}
\email{andrea.fortini@uni-bayreuth.de}
\affiliation{Theoretische Physik II, Physikalisches Institut, Universit\"at Bayreuth, Universit\"atsstra{\ss}e 30, D-95447 Bayreuth,
Germany}
\pacs{61.20.Ja,83.80.Hj,81.16.Dn}

\begin{abstract}
A  mixture of hard-sphere particles and model emulsion droplets is studied with a Brownian dynamics simulation. We find that the addition of non-wetting emulsion droplets to a suspension of pure hard spheres can lead to both gas-liquid and fluid-solid phase separations. Furthermore, we find  a stable fluid of hard-sphere clusters. The stability is due to the  saturation of the attraction that occurs when the surface of the droplets is completely covered with colloidal particles.  At larger emulsion droplet densities a percolation transition is observed. The resulting networks of colloidal particles show dynamical and mechanical properties typical of a colloidal gel. The results of the model are in good qualitative agreement with recent experimental findings [Koos, E. and Willenbacher, N. (2011). Science, 331, p.897] in a mixture of colloidal particles and two immiscible fluids.
\end{abstract}

\maketitle

Controlling the flow and the mechanical properties of complex fluids has important implications in both science and technology. Therefore, the possibility to change the viscoelastic and flow properties of  colloidal or nanoparticle suspensions simply by adding a small amount water  is intriguing in its simplicity and elegance. This surprising result was obtained experimentally by~\citet{Koos:2011hf} (see also Ref.~\cite{Butt:2011hf}). They found that the addition of an immiscible non-wetting fluid to a stable colloidal suspension leads to the formation of emulsion droplets that act as a bonding agent between the colloidal particles. As observed in microscopy images, the transition from a fluid to a gel is driven by the formation of a network of colloidal particles bonded by emulsion droplet. 
It is well known that colloids trapped at droplet interfaces reduce the interfacial free energy  (the Pickering effect~\cite{Pieranski1980}), and this basic physical mechanism is used
 to produce colloidal clusters\cite{Manoharan2003,Lauga2004,Schwarz:2011} as well as to stabilize emulsions~\cite{Sacanna:2007,Jansen2011} and bigels~\cite{Sanz:2009gt}. 
 
A detailed understanding of the gel-transition mechanism driven by a low density of secondary fluid emulsion droplets is  missing at present. The aim of this article is to introduce a minimal model that describes the network formation when non-wetting emulsion droplets are mixed with a stable suspension of purely repulsive colloidal particles. 

We find that  the Pickering effect drives the cluster-cluster aggregation that is behind a transition from a fluid of clusters to a percolated network, and that the properties of these networks are those of colloidal gels~\cite{Lu:2008fo,Lodge:1997uk,Zaccarelli:2007hp,Eberle:2011fi}. Since the Pickering  mechanism saturates once the droplet surfaces are completely covered with colloids, a fluid of stable clusters is achievable without the addition of a long range repulsion~\cite{Campbell2005,Lu2006}.
In addition we find that the Pickering effect can induce  gas-liquid and gas-solid phase separations. 

We  describe the suspension  as a binary mixture of $N_{\textrm c}$ colloidal particles with hard-sphere 
diameter $\sigma_{\textrm c}$ and $N_{\textrm d}$ spherical droplets of diameter
$\sigma_{\textrm d}$.  The total interaction energy $U$ is the sum of
colloid-colloid, droplet-droplet and colloid-droplet interactions,
\begin{eqnarray}
\frac{U}{k_{\rm B}T} &=&\sum_{i<j}^{N_{\textrm c}} \phi_{cc}(|\vec r_{i}-\vec r_{j}|
)+\sum_{i<j}^{N_{\textrm d}} \phi_{dd}(|\vec R_{i}-\vec R_{j}| ) \nonumber
\\ && + \indent \sum_{i}^{N_{\textrm c}} \sum_{j}^{N_{\textrm d}} \phi_{cd}(|\vec
r_{i}-\vec R_{j}|),
\end{eqnarray}
where $k_{\rm B}$ is the Boltzmann constant, $T$ is the temperature, $\vec{r}_i$ is the center-of-mass position of colloid $i$, $\vec
R_{j}$ is the center-of-mass position of droplet $j$, $\phi_{cc}$ is
the colloid-colloid pair interaction, $\phi_{cd}$ is the
colloid-droplet pair interaction, and $\phi_{dd}$ is the
droplet-droplet pair interaction.

The colloid-colloid and droplet-droplet interactions are hard-sphere
like
\begin{equation} 
 \phi_{  ii }(r)=\left \{ 
\begin{array}{ll}
 - 1 +  \left( \frac{\sigma_i}{r} \right)^{36} &  r<  \sigma_i\\
0
    &  \text{otherwise,} \\
\end{array} \right . 
\label{eq:hs}
\end{equation}
where $i=d$, for the droplet-droplet interaction and $i=c$ for
colloid-colloid interaction.  The steep repulsive continuos potential was chosen to allow integration with a Brownian Dynamics algorithm~\cite{Allen1987}.
The repulsion between droplets is aimed at modeling the repulsion of 
charged droplets; with this assumption droplets coalescence can also be neglected.
Furthermore, the shape of the droplets remains spherical.
  
The colloid-droplet interaction is aimed at modeling the Pickering effect. The loss of interfacial
energy~\cite{Pieranski1980} when a particle is trapped at the surface
of a droplet is modeled simply by a parabolic  well of depth $\epsilon$
\begin{equation} 
\phi_{\textrm cd}(r)= \left \{ 
\begin{array}{ll}
 \frac{1}{k_{\rm B} T} [ a (r-B)^{2}+c]  & r<  \frac{\sigma_d+\sigma_{\textrm c} }{2}\\
    0 & \mbox{otherwise,}
\end{array} \right . 
\label{eq:DC}
\end{equation}
where a=$(-\epsilon + \sqrt{\epsilon^{2}+r_{0}^{2}})/2 r_{0}^{2}$, c=$ (\epsilon - \sqrt{\epsilon^{2}+r_{0}^{2}})/2)$, and $B= (\sigma_d+\sigma_{\textrm c})/2-r_{0}$. 
 The value $r_{0}=0.1$ determines the contact angle
between droplets and colloids to $\Theta=150^{\circ}$, and 138$^{\circ}$, for
$\sigma_{\textrm d}/\sigma_{\textrm c}$=1.5, and 0.75, respectively. 
Curvature and merging of droplets are neglected. A more detailed  model~\cite{Jansen2011} would be necessary to describe all possible droplets behaviors.

We carried out  Brownian Dynamics (BD) simulations~\cite{Allen1987} (hydrodynamics interactions are neglected) with time step$\delta t =1 \times 10^{-5}
\tau_{\textrm B}$, with $\tau_{\textrm B}=\sigma_c^2/D_{\textrm 0}$, where $D_{0}$ is the
Stokes-Einstein diffusion coefficient. The random forces mimic the
interaction between particles and solvent, and are sampled from a
Gaussian distribution with variance $2 D_{0} \delta t$. The system is at
constant temperature. 
The BD simulations were carried out for $N_{\textrm c}$=1113
hard spheres, at fixed packing fraction $\phi_{\textrm c}$=0.1, and for a number
of droplets in the range $N_{\textrm d}$=10 to 667, i.e. at
different droplets volume fractions $\phi_{\textrm d}$. To improve the
statistical accuracy of the sampled quantities each state point was simulated four or eight
times. 

\begin{figure}
\includegraphics[width=8cm]{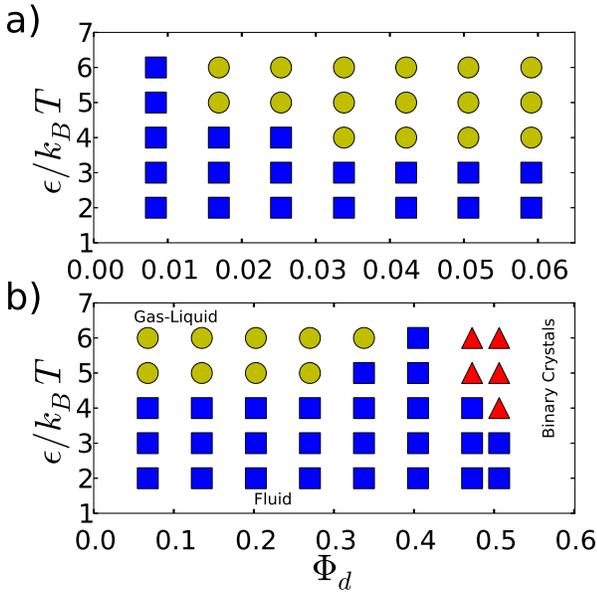}
\caption{(Color online) Equilibrium phases at $\phi_{c}=0.1$. Squares indicate homogeneous fluid phases, circles indicate gas-liquid coexistence, while triangles indicate fluid-binary crystal coexistence.  a) Droplet size $\sigma_{\textrm d}/\sigma_{\textrm c}$=0.75. b)Droplet size $\sigma_{\textrm d}/\sigma_{\textrm c}$=1.5.}
\label{fig:phd}
\end{figure}

The equilibrium phases of the model are investigated by
carrying out simulations (for $t=1000 \, \tau_{\textrm B}$) at fixed $\phi_{c}=0.1$ and  different values of  $(\phi_{d},\epsilon)$. The final phase for each state point is reported in Fig.~\ref{fig:phd}.
For both $\sigma_{\textrm d}/\sigma_{\textrm c}$=0.75, and 1.5 we find a region where the fluid is stable (squares) and a region where gas-liquid phase separation occurs (circles). We stress that this separation is not  demixing of two species, but separation in a fluid that is rich in both droplets and colloids, and a dilute gas.  Additionally, for $\sigma_{\textrm d}/\sigma_{\textrm c}$=1.5 we find a region where the fluid separates in a fluid and a binary crystal. In Fig.\ref{fig:phd}b) a critical point and a triple point are recognizable.

\begin{figure}
\includegraphics[width=8cm]{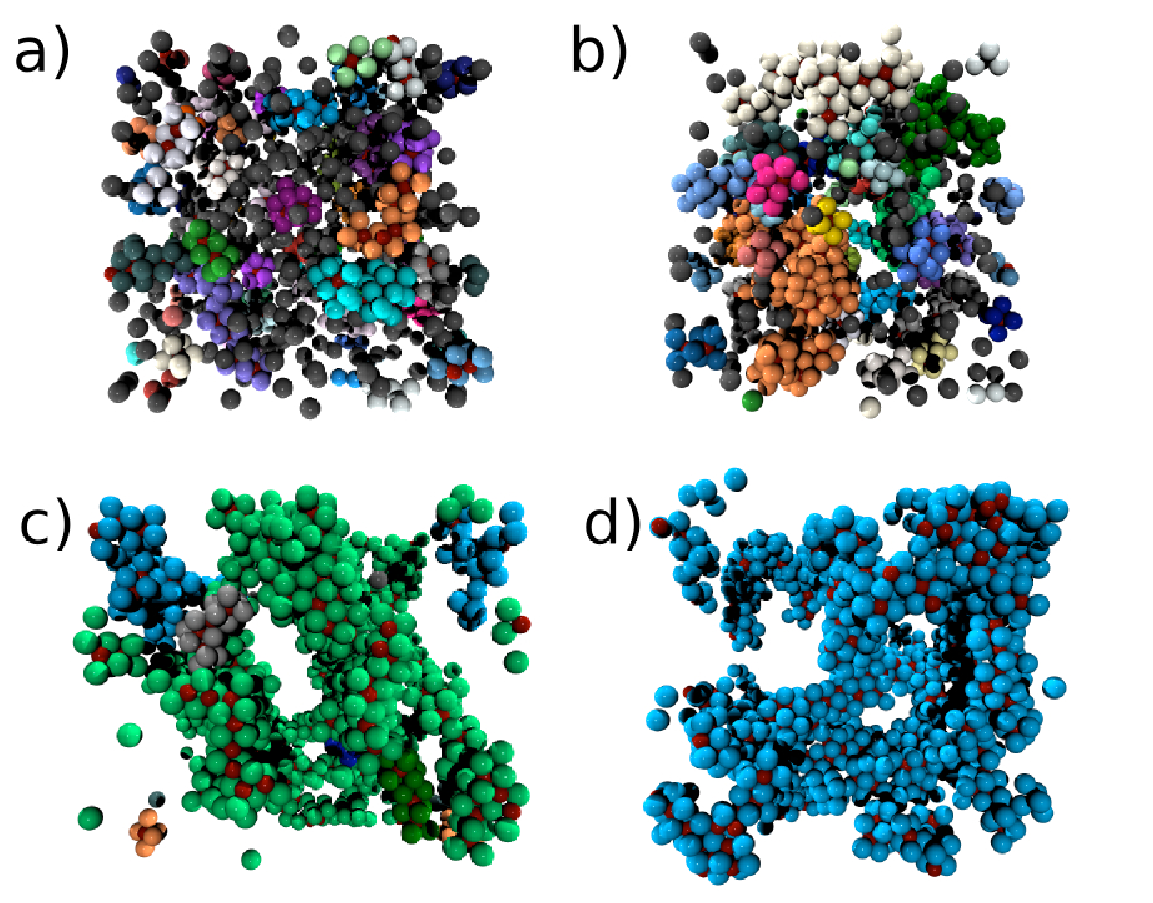}
\caption{(Color online) Simulation snapshots at  $t=800 \, \tau_{\textrm B}$
  for $\sigma_{\textrm d}=0.75$. Droplets are shown as small red (dark grey) spheres while  different clusters are colored (shaded) differently. a) Fluid of clusters
  at $\Phi_{\textrm d}$=0.63\% b) Fluid of clusters at $\Phi_{\textrm d}$=1.05\% c) Fluid at
  percolation point $\Phi_{\textrm d}$=1.48\% d)$\Phi_{\textrm d}$=1.9\% Percolated
  state.}
\label{fig:snaps}
\end{figure}

To study the cluster and gelation transitions we choose state points within the  gas-liquid envelope (see Fig. 8 of Ref.~\cite{Fortini:2008cx}). The  energy scale of the droplet-particle interaction was fixed at $\epsilon=100 k_{\textrm B} T $. 
Fig.~\ref {fig:snaps} shows simulation snapshots at different droplets
volume fractions $\phi_{\textrm d}$, at $t=800 \, \tau_{\textrm B}$. 
Different clusters are colored (shaded) differently.  A cluster is defined as a group of bonded particles. A bond between two particles is formed only when they share the surface of the same droplet. 
At low droplet volume fractions, as shown in Figs.~\ref{fig:snaps}a)-b), many  clusters  coexists with single particles and the system is therefore characterized by a high degree of dynamical heterogeneity.  
At the higher droplet volume fractions (Figs.~\ref{fig:snaps}c)-d)) a large cluster spans the entire simulation box. 
A visual inspection of the  snapshots indicates that a transition from a fluid of  clusters to a percolated network occurs, and that it is controlled by the amount of emulsion droplets. 

\begin{figure}
\includegraphics[width=8cm]{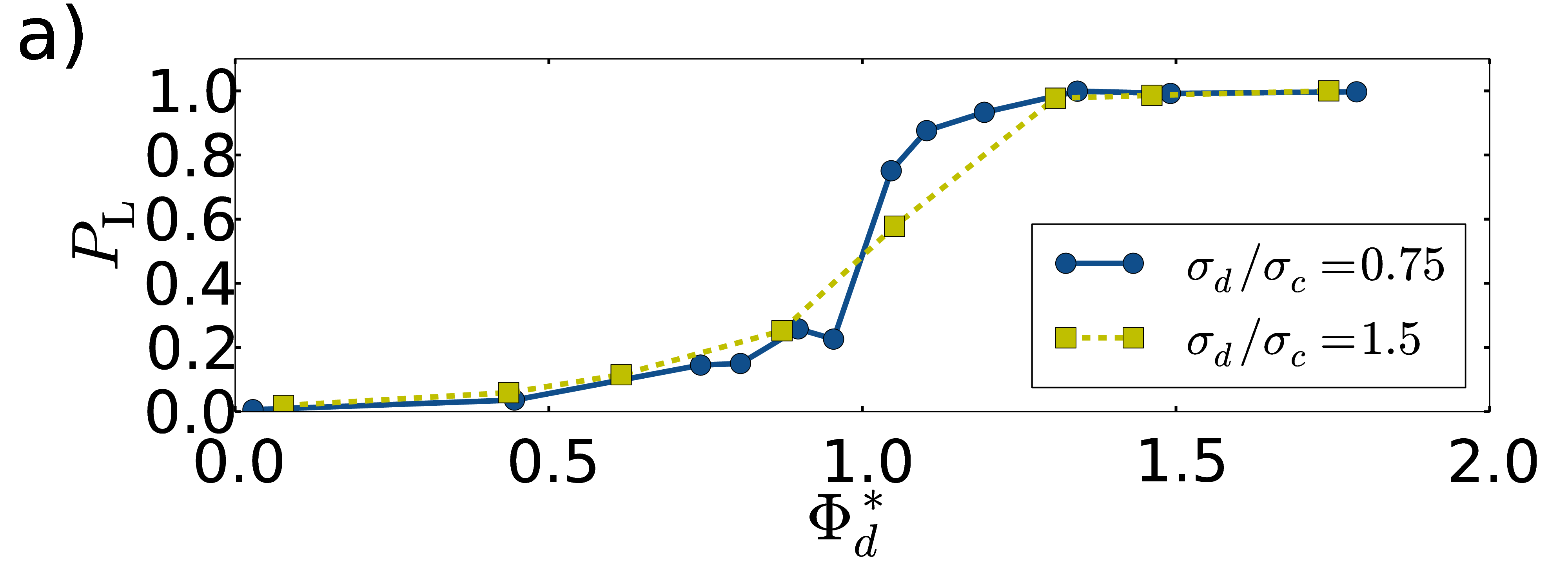}
\includegraphics[width=8cm]{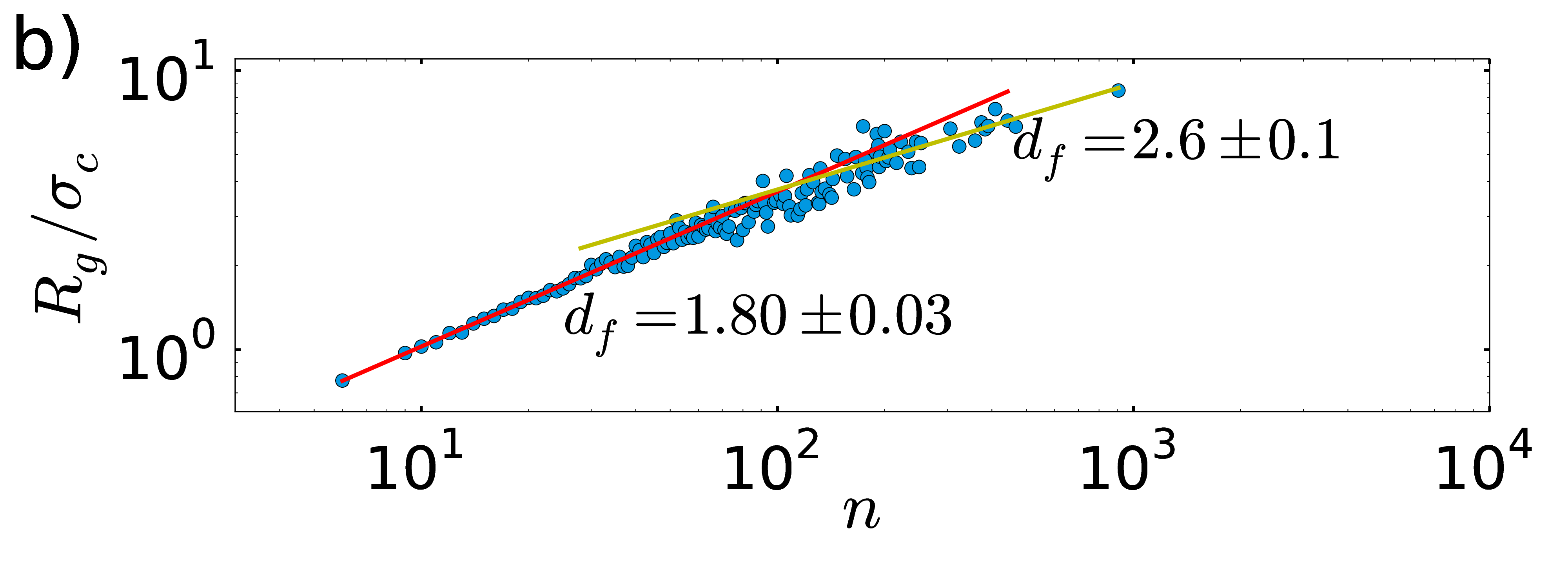}
\includegraphics[width=8cm]{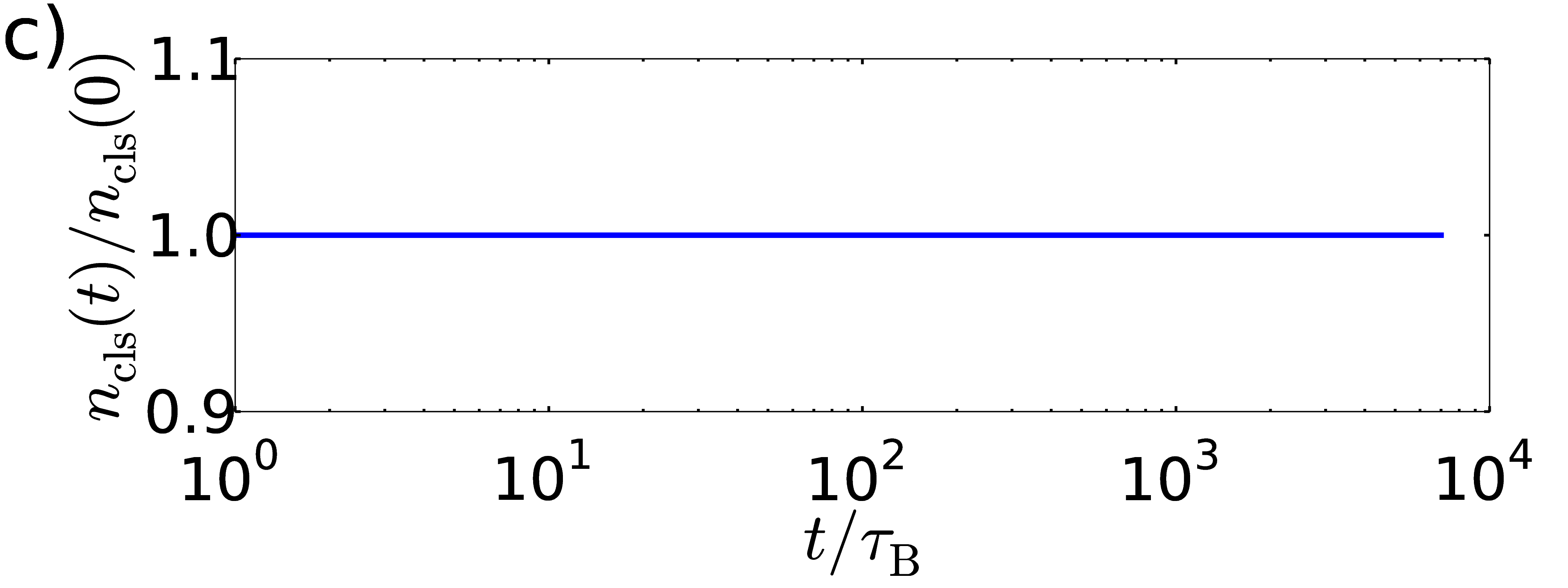}
\caption{(Color online) a) The probability $P_{\textrm L}$ that a particle
  belongs to the largest cluster as a function of the rescaled droplet
  volume fractions $\phi_{\textrm d}^{*}=\phi_{\textrm d}/\phi_{\textrm d}^{c}$. The critical
  values of the droplet volume fraction are $\phi_{\textrm d}^{c}$=0.01415, 0.039,
  for droplet sizes $\sigma_{\textrm d}/\sigma_{\textrm c}$=0.75, and 1.5,
  respectively. b) The radius of gyration of the clusters $R_{\textrm g}$
  versus the size $n$ of the cluster. c) The time evolution of the number of clusters $n_{\rm cls}(t)/n_{\rm cls}(0)$ for  $\phi_{\textrm d}^{*}=0.75$. The number of clusters at $t=0$ is $n_{\rm cls}(0)$=48.}
\label{fig:perc}
\end{figure}

The percolation transition was analyzed by carrying out a  cluster analysis of eight independent runs at different  droplet volume fractions $\phi_{\textrm d}$.  Figure~\ref{fig:perc}a) shows the
probability $P_{\textrm L}$ that a particle belongs to the largest cluster as a function of the rescaled droplet volume fractions $\phi_{\textrm d}^{*}=\phi_{\textrm d}/\phi_{\textrm d}^{c}$.
The critical volume fraction $\phi_{\textrm d}^{c}$ is defined as the volume fraction of the percolation transition, i.e. where $P_{\textrm L}=0.5$.
At small $\phi_{\textrm d}^{*}$ we observe a fluid of increasingly large clusters, until a sharp transition to a percolated network is observed at large $\phi_{\textrm d}^{*}$.
The percolation critical values  are $\phi_{\textrm d}^{c}$=0.01415, 0.039, for droplet sizes
$\sigma_{\textrm d}/\sigma_{\textrm c}$=0.75, and 1.5, respectively.  The small
amounts of secondary fluid necessary for the percolation transition to occur are in good qualitative agreement with the results of~\citet{Koos:2011hf} who found a transition from a fluid to a gel behavior at a volume fraction of water $\phi=0.003$ in a suspension of particles with volume fraction $\phi=0.111$.  
We find that the transition value depends on the relative size of droplets and particles. Therefore we expect an influence of the droplets polydispersity on the location of the percolation transition.
More information about the percolation mechanism can be extracted from the analysis (Fig.~\ref{fig:perc}b))  of the radius of gyration  $R_{\textrm g}=\frac{1}{n} \sqrt{\sum_{i<j}|r_{i}-r_{j}|^{2}}$ of  clusters of size  $n$ (number of constituents particles). 
For small sizes  the clusters grow with a fractal dimension $d_{\textrm f}=1.80 \pm 0.05$;  this value suggests  three-dimensional diffusion limited cluster aggregation (DLCA)~\cite{Meakin:1983,Kolb:1983,Rottereau:2004}. At larger cluster sizes the  aggregation is characterized by a fractal dimension $d_{\textrm f}=2.6 \pm 0.1$, i.e. a random aggregation of different mesoscopic clusters, which form the different branches of 
the final percolating cluster.  This value of $d_{\textrm f}$ is also consistent with the fractal dimension $d_{\textrm f}=2.7 \pm 0.1$ obtained from the slope of the cumulative integral of the radial distribution function of the percolated networks. Since, the crossover to random aggregation occurs at sizes that are about a quarter the lateral size of the simulation box, we do not expect finite size effects to dramatically change the location of the crossover.
Lastly, in Fig.~\ref{fig:perc}c) the time evolution of the number of clusters $n_{\rm cls}(t) $ is shown. Surprisingly, the number of clusters remains constant during the simulation time, indicating that the fluid of clusters is stable. 

\begin{figure}
\includegraphics[width=8cm]{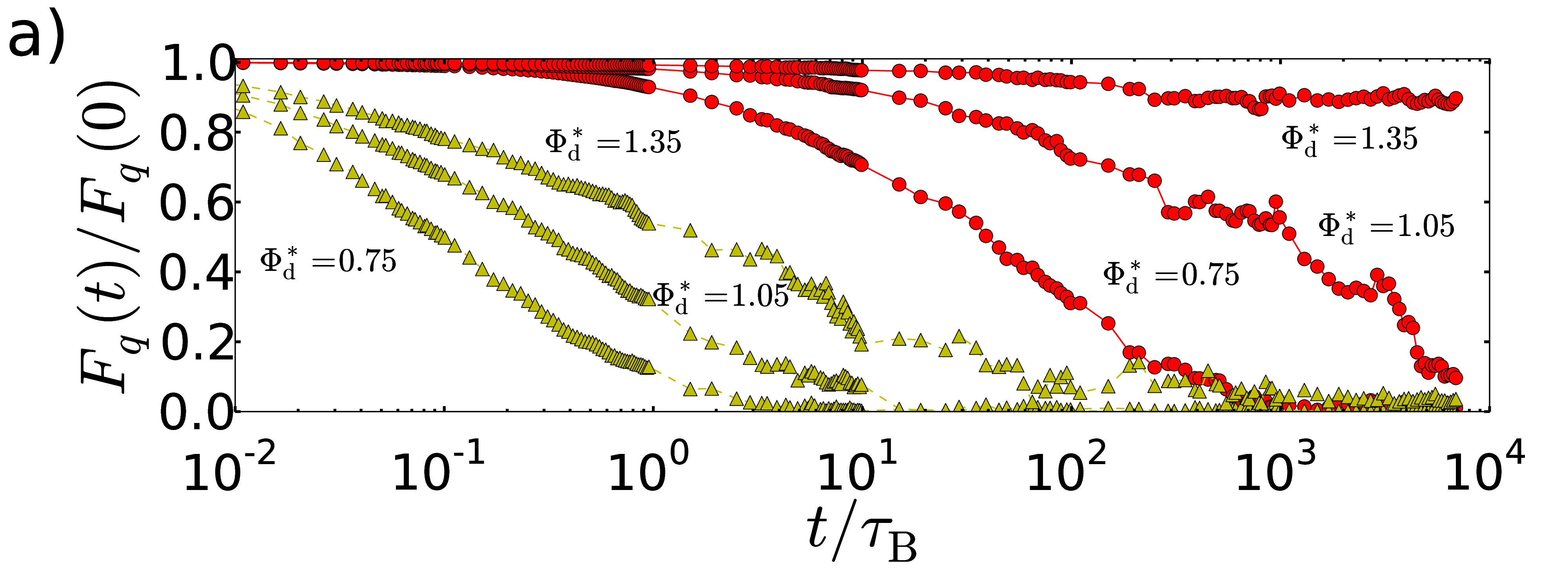}
\includegraphics[width=8cm]{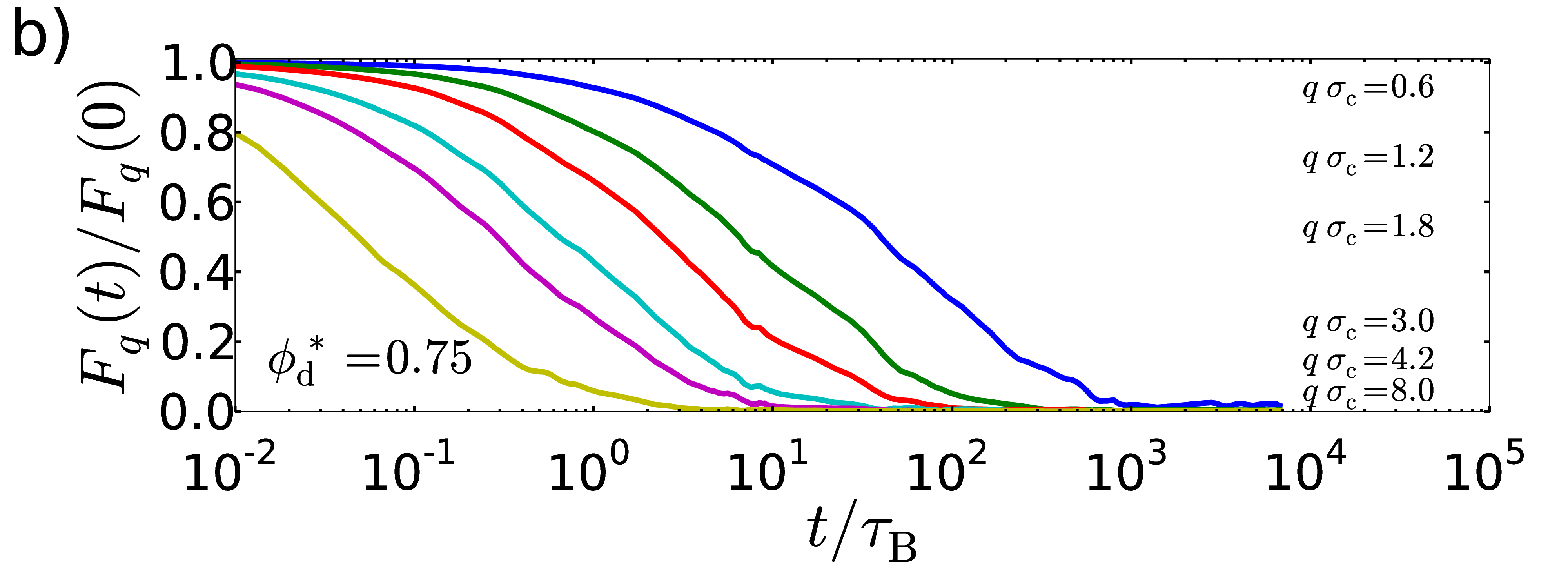}
\includegraphics[width=8cm]{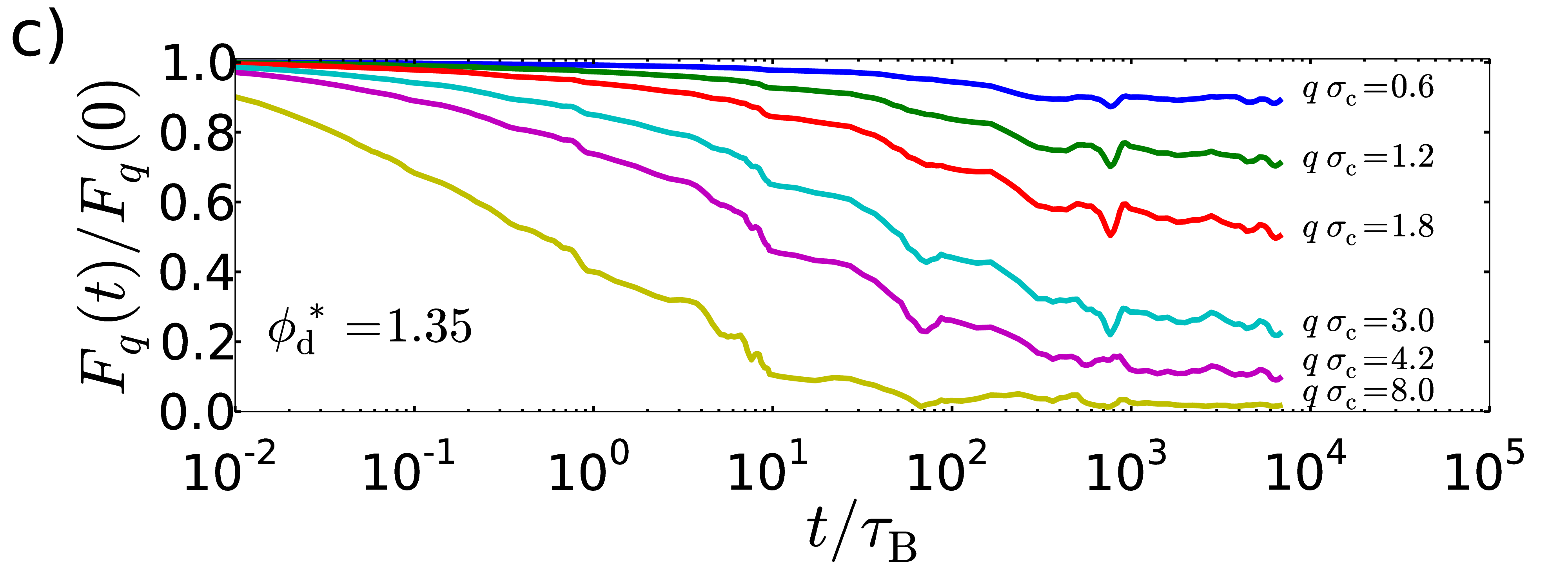}
\caption{(Color online) Self-intermediate scattering function $F(q,t)$
  as a function of time $t$. a) For $q\sigma_{\textrm c}=6.28$ (yellow curves)
  and $q\sigma_{\textrm c}=0.6$ (red curves) for scaled droplet volume
  fractions $\phi_{\textrm d}^{*}$=0.75, 1.05, 1.35.  b) For $\phi_{\textrm d}^{*}$=0.75,
  and $q \sigma_{\textrm c}$=0.6, 1.2, 1.8, 3.8, 4.2, and 8.0, from top to
  bottom.  c) Same as b), but for  $\phi_{\textrm d}^{*}$=1.35. }
\label{fig:fk}
\end{figure}

We study dynamics via the colloid self-intermediate scattering function
\begin{equation} 
F_q(t)=\frac{1}{N_{c}} \langle   \sum_{j}^{N_{c}} e^{-i \vec q \cdot \vec r_{j}(t)} e^{i \vec q \cdot \vec r_{j}(0)}\rangle \ ,
\label{eq:self}
\end{equation}
where $\vec r_{j}(t)$ is the position vector of particle $j$ at time $t$, and  $\vec q$ is a wave vector.
The brackets indicate an average over time origins and four independent simulations. The sampling was done for a time $t/\tau_{\rm B}=6000$.
Fig.~\ref{fig:fk}a) shows the behavior of $F_q(t)$ as a function of $t$ at fixed
$q\sigma_{\textrm c}=6.28$ (yellow triangles) and $q\sigma_{\textrm c}=0.6$ (red circles)
for different state points.  
For $q \sigma_{\textrm c}=6.28$, $\phi_{\textrm d}^{*}$=0.75, and 1.05, the function decays to zero. This behavior represents the normal dynamics of a fluid, where particles  quickly become uncorrelated (in time) at length scales of the order of a particle diameter. At $\phi_{\textrm d}^{*}$=1.35, i.e. well inside the
percolation region, the decay is much slower. At $q \sigma_{\textrm c}=0.6$ (red circles) the transition from  ergodic  $\phi_{\textrm d}^{*}$=0.75 to  non-ergodic behavior  at $\phi_{\textrm d}^{*}$=1.35 is a strong indication  that a structural arrest  characteristic of a gel transition~\cite{Zaccarelli:2007hp} has occurred at percolation.
Figure~\ref{fig:fk}b) shows that all wave vectors remain ergodic at $\phi_{\textrm d}^{*}$=0.75. Therefore, the reason for the stability of the fluid of clusters (Fig.~\ref{fig:perc}c)) is not slow dynamics, but the saturation of the attraction. 
The  structural arrest at $\phi_{\textrm d}^{*}$=1.35  is  clearly shown in Fig.~\ref{fig:fk}c), with a plateau (non-ergodic behavior) that develops fully only for small $q \sigma_{\textrm c}$.

\begin{figure}
\includegraphics[width=8cm]{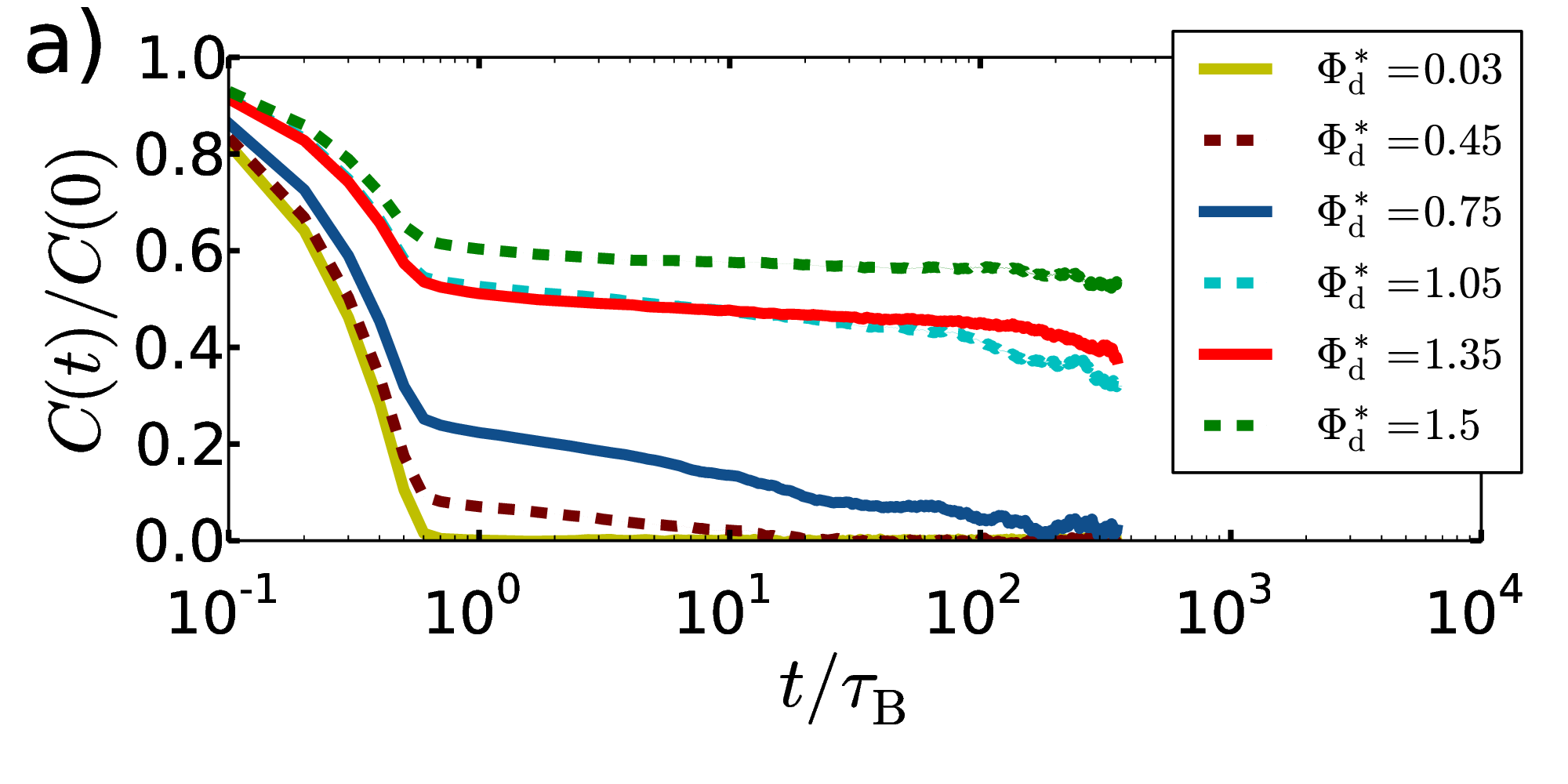}
\includegraphics[width=8cm]{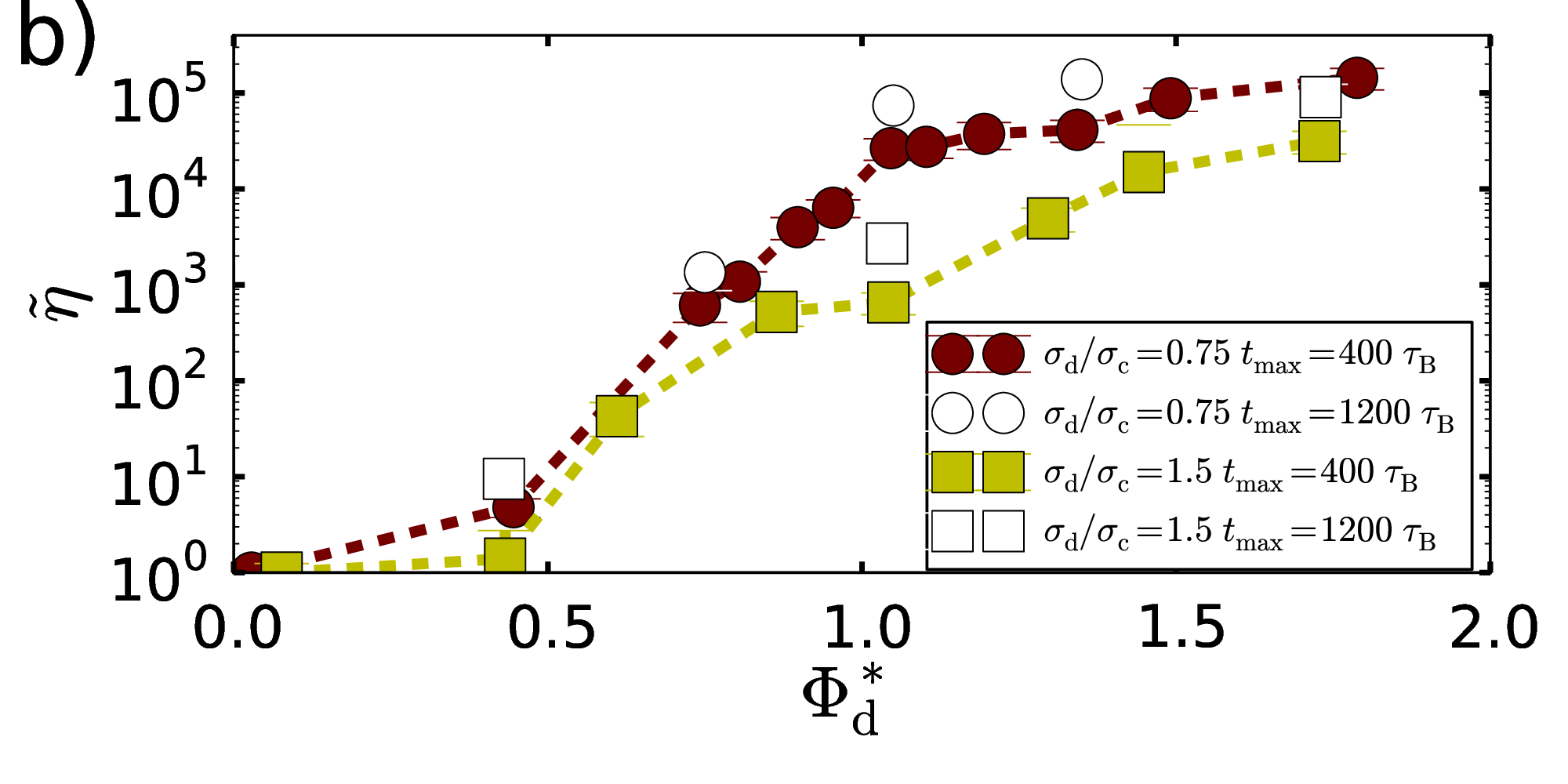}
\caption{(Color online) a) Stress autocorrelation function $c_{s}(t)/c_{s}(0) $
  at $\sigma_{\textrm d}/\sigma_{\textrm c}$=0.75 and different values of scaled droplet volume fraction $\phi_{\textrm d}^{*}$
  b) Scaled viscosity $\tilde \eta =\eta(t_{\textrm max})/\eta(t_{\textrm max},\phi_{\textrm d}^{*}=0.03)$ for two different droplet sizes
  $\sigma_{\textrm d}/\sigma_{\textrm c}$=0.75, and 1.5.}
\label{fig:str}
\end{figure}
The mechanical response of the system is characterized via the shear stress autocorrelation function $c_{s}(t)$
\begin{equation}
c_{s}(t)=\frac{V}{k_{\textrm B} T} \langle S_{\alpha \beta}(0) S_{\alpha \beta}(t) \rangle
\label{eq:auto}
\end{equation}
where V is the system volume and the brackets indicate an average over
time origins and eight independent simulations. The sampling was done for  $t/\tau_{\rm B}=400$ and  $t/\tau_{\rm B}=1200$. 
The function $S_{\alpha \beta}(t)$ represents the off-diagonal elements of the stress tensor
\begin{equation} 
S_{\alpha \beta}(t)=\frac{1}{V} \sum_{i}^{N-1}\sum_{j=i+1}^{N}  r^{ij}_{\alpha }(t)  f^{ij}(t)_{\beta} \ ,
\label{eq:stress}
\end{equation}
where $r^{ij}_{\alpha }$ is the $\alpha$-component of the distance vector $\vec r_{ij}$ between particles $i$, and $j$, and $f^{ij}_{\beta}$ is the $\beta$-component of the pairwise force. 
Fig.~\ref{fig:str}a) shows the stress autocorrelation function as a function of time.  In a fluid, ($\phi_{\textrm d}^{*}=0.03$) $c_{s}(t)$ quickly decays to zero with a stretched exponential form. The transition to a power law decay (linear region in the log-log plot for $\phi_{\textrm d}^{*}=0.45, 0.75$) is  a sign of an approaching gel transition~\cite{Lodge:1997uk}. When a percolated network is formed $(\phi_{\textrm d}^{*}=1.05, 1.35, 1.5$) a plateau in the correlation function develops. The plateau is an indication that the system is able to sustain stress.  
At equilibrium, the time integral of  $c_{s}(t)$ gives the fluid viscosity via the Green-Kubo relation~\cite{Hansen1986}
\begin{equation}
\eta=\lim_{t_{\textrm max}\rightarrow \infty} \eta(t_{\textrm max}) = \lim_{t_{\textrm max}\rightarrow \infty} \int _{0}^{t_{\textrm max}} c_{s}(t) dt,
\label{eq:eta}
\end{equation}
In Fig.~\ref{fig:str}b) the values of $\eta(t_{\textrm max})$ are plotted against the scaled droplet volume fraction $\phi_{\textrm d}^{*}$ for two different droplet sizes
$\sigma_{\textrm d}/\sigma_{\textrm c}$=0.75, and 1.5.  
The transition from liquid-like to solid-like behavior is not sharp because of the finite time used in the simulation. We stress that the values plotted in Fig.~\ref{fig:str}b) represent the viscosity only if the stress correlation function decays to zero. Nonetheless, the value of the integral (\ref{eq:eta}) is an indicator of  the changing mechanical response of the system and it shows that  small droplets influence the flow behavior more than large droplets. 

In this article a model was introduced that describes the formation of a fluid of stable clusters and gels when emulsion droplets are added to a  hard spheres suspension.
The particle aggregation into clusters is due to the Pickering effect, while the mechanism for cluster-cluster aggregation is due to colloids bonding to multiple droplets. 
The cluster-cluster aggregation leads to the formation of a spanning network of colloidal particles via a percolation transition. A crossover from DLCA to random aggregation of  large clusters is observed, and the critical percolation density of emulsion droplets depends on the droplet to particle diameter ratio. 
The analysis of the dynamics via the colloid self-intermediate scattering function shows a strong structural arrest at small wavelengths, indicating the occurrence of a gel transition in correspondence of the percolation transition.  
The analysis of the stress correlation function also suggests the ability of the percolated networks to sustain stress. Therefore, both the dynamical and mechanical properties of these networks are those of gels, in agreement with the experiments~\cite{Koos:2011hf}.
Due to bond saturation there is a lack of available attraction in the final configurations that, combined with  a strong particles-droplet  bonding, suggests that aggregation and collapse of clusters 
and gels is hindered even at very long timescales.  

Furthermore, we traced the stability phase diagram for the model and found that both gas-liquid and  fluid-solid separations are possible. In particular, we found that binary crystals can self-assemble at high concentration of emulsion droplets.
These phases were not yet observed experimentally in Pickering emulsions. Although their realization represent a big technical challenge, emulsions could prove to be a novel way to control the formation of complex colloidal crystal structures. 

\begin{acknowledgments}  
I would like to thank Thomas Fischer and Matthias Schmidt for discussions. 
I acknowledge the DFG for support via SFB840/A3.
\end{acknowledgments}

\end{document}